%% file: template-epjc.tex
\pdfoutput=1
\RequirePackage{fix-cm}
\documentclass[twocolumn,epjc3]{svjour3}  
\smartqed  
\RequirePackage{graphicx}
\RequirePackage{subfigure}
\RequirePackage{mathptmx}      
\RequirePackage{flushend}
\RequirePackage{latexsym}
\RequirePackage[numbers,sort&compress]{natbib}
\RequirePackage[colorlinks,citecolor=blue,urlcolor=blue,linkcolor=blue]{hyperref}
\RequirePackage{xspace}
\RequirePackage{tikz}
\RequirePackage{xcolor}
\definecolor{lime}{HTML}{A6CE39}
\DeclareRobustCommand{\orcidicon}{
	\begin{tikzpicture}
	\draw[lime, fill=lime] (0,0) 
	circle [radius=0.16] 
	node[white] {{\fontfamily{qag}\selectfont \tiny ID}};
	\draw[white, fill=white] (-0.0625,0.095) 
	circle [radius=0.007];
	\end{tikzpicture}
	\hspace{-2mm}
}
\foreach \x in {A, ..., Z}{\expandafter\xdef\csname orcid\x\endcsname{\noexpand\href{https://orcid.org/\csname orcidauthor\x\endcsname}
			{\noexpand\orcidicon}}
}
\newcommand*{\selectron}{\ensuremath{\tilde{e}}\xspace}
\newcommand*{\smu}{\ensuremath{\tilde{\mu}}\xspace}

\newcommand*{\stau}{\ensuremath{\tilde{\tau}}\xspace}

\newcommand*{\ninoone}{\ensuremath{\mathchoice%
      {\displaystyle\raise.4ex\hbox{\(\displaystyle\tilde\chi^0_1\)}}%
         {\textstyle\raise.4ex\hbox{\(\textstyle\tilde\chi^0_1\)}}%
       {\scriptstyle\raise.3ex\hbox{\(\scriptstyle\tilde\chi^0_1\)}}%
 {\scriptscriptstyle\raise.3ex\hbox{\(\scriptscriptstyle\tilde\chi^0_1\)}}}\xspace}
\journalname{Eur. Phys. J. C}
\begin{document}

\title{Prospects for slepton pair production in the future $e^{-}e^{+}$ Higgs factories
}
\author{\orcidA{}Jiarong Yuan\thanksref{addr1,addr2}
        \and
        \orcidB{}Huajie Cheng\thanksref{addr1,addr3} 
        \and
        \orcidC{}Xuai Zhuang\thanksref{e1,addr1}
}

\thankstext{e1}{e-mail: zhuangxa@ihep.ac.cn(corresponding author)}
\institute{Institute of High Energy Physics, Chinese Academy of Science, Yuquan Road 19B, Shijingshan District, Beijing 100049, China \label{addr1}
           \and
           University of the Chinese Academy of Sciences, Yuquan Road 19A, Shijingshan District, Beijing 100049, China \label{addr2}
           \and
           Department of Applied Physics, Naval University of Engineering, Jiefang Blvd 717, Qiaokou District, Wuhan 430033, China \label{addr3}
}
\maketitle
\begin{abstract}
The Circular Electron Positron Collider (CEPC) with a center-of-mass energy $\sqrt{s}$ = 240 GeV is proposed to serve as a Higgs factory, while it can also provide good opportunity for new physics searches at lower energy, which are difficult in hadron colliders but well-motivated by some theory models such as dark matter.
This paper presents the sensitivity study of direct stau / smuon production at CEPC using full Monte Carlo (MC) simulation. 
With the assumption of flat 5\% systematic uncertainty, the CEPC has the potential to discover the production of combined left-handed and right-handed stau up to 116 GeV if exists, or up to 113 GeV for the production of pure left-handed / right-handed stau; the discovery potential of direct smuon reaches up to 117 GeV with the same assumption. 
Due to the conserved assumption of systematic uncertainty and limited reliance on the reconstruction and detector geometry in this study, the results can be used as reference for similar searches in other electron positron colliders at a center-of-mass energy close to 240 GeV, such as Future Circular Collider $e^{+}e^{-}$ (FCC-ee) and the International Linear Collider (ILC). 
\end{abstract}

\section*{Declarations}
\label{sec:declarations}
\input{tex/declarations.tex}

\section{Introduction}
\label{sec:intro}
\input{tex/introduction.tex}

\section{Detector, Software and Samples}
\label{sec:samples}
\input{tex/samples_software.tex}

\section{Search for the direct slepton production}
\label{sec:searchdl}
\input{tex/searchdl.tex}

\pagebreak
\clearpage

\section{Conclusion}
\label{sec:conclusion}
\input{tex/conclusion.tex}

\section{Acknowledgments}
\label{sec:acknowledgments}
\input{tex/acknowledgments.tex}


\bibliographystyle{spphys}       
\bibliography{DSL.bib}
\end{document}

%% file: tex/declarations.tex
\subsection*{Funding}
This study was supported by the National Key Programme (Grant NO.: 2018YFA0404000).
\subsection*{Availability of data and material}
The data used in this study won't be deposited, because this study is a simulation study without any experiment data.

%% file: tex/introduction.tex
Supersymmetry (SUSY)~\cite{Golfand:1971iw,Volkov:1973ix,Wess:1974tw,Wess:1974jb,Ferrara:1974pu,Salam:1974ig,Martin:1997ns} proposes that there is a superpartner, known as sparticle, for every Standard Model (SM) particle, whose spins are different by a half from the corresponding SM particle.
With $R$-parity~\cite{Farrar:1978xj} conserved, SUSY particles are produced in pairs, and the lightest supersymmetric particle (LSP) is stable, which is a dark matter candidate~\cite{Goldberg:1983nd,Ellis:1983ew}. 

The linear superpositions of charged and neutral Higgs bosons and electroweak gauge bosons formed two charged mass eigenstates called charginos and four neutral mass eigenstates called neutralinos. 
The superpartner of a lepton is a slepton whose chirality is the same as the lepton's chirality.
The slepton mass eigenstates formed from superpositions of left-handed sleptons and right-handed sleptons.

Models with light sleptons satisfies the dark matter relic density measurements~\cite{Vasquez:2011}.
And light sleptons can take part in the coannihilation of neutralinos~\cite{Belanger:2004ag,King:2007vh}.
Models with light smuons can explain $(g-2)_{\mu}$ excess~\cite{Endo:2019bcj}.
In gauge-mediated~\cite{Dine:1981gu,AlvarezGaume:1981wy,Nappi:1982hm} and anomaly-mediated~\cite{Randall:1998uk,Giudice:1998xp} SUSY breaking models, the mass of sleptons are expected to be of the order of magnitude of 100 GeV.

Stau (smuon) mass below 86 - 96 (95 - 99) GeV are excluded for mass difference between \stau (\smu) and \ninoone larger than 7 (4) GeV~\cite{LEPslepton,Heister:2001nk,Heister:2003zk,Abdallah:2003xe,Achard:2003ge,Abbiendi:2003ji}.
Using 139 fb$^{-1}$ of data collected by ATLAS, sleptons (\selectron or \smu) are constrained to have masses above 251 GeV for a mass splitting of 10 GeV, with constraints extending down to mass splittings of 550 MeV at the LEP slepton limits (73 GeV)~\cite{SUSY-2018-16}.
With a massless LSP, for slepton pair production (\selectron or \smu) masses up to 700 GeV are excluded using 139 fb$^{-1}$ of data collected by ATLAS assuming three generations of mass-degenerate sleptons~\cite{SUSY-2018-32}.
With a massless LSP, for direct smuon production masses up to 310 GeV are excluded using 35.9 fb$^{-1}$ of data collected by CMS~\cite{CMS-SUS-17-009}.
With a massless LSP, for direct stau production masses from 120 GeV to 390 GeV are excluded using 139 fb$^{-1}$ of data collected by ATLAS~\cite{SUSY-2018-04}.
In a degenerate production model, in which both left- and right-handed \stau pairs are produced, with a massless LSP \stau masses are excluded up to 150 GeV using 77.2 fb$^{-1}$ collected by CMS~\cite{CMS-SUS-18-006}. 
However, for the cases with massive LSP, especially when the mass split of slepton and LSP is very small, the sensitivity from LHC is limited.

CEPC has higher center-of-mass energy than LEP.
Compared to LHC, CEPC has very clean collision environment, which means less backgrounds. 
Besides, reconstruction and identification efficiencies for tracks and single particles (e.g. muon) at CEPC are higher than LHC, which ensures sufficient sensitivities for the scenarios with very soft objects~\cite{Ruan:2018yrh}.

Just like CEPC, ILC and FCC-ee are peoposed electron positron colliders~\cite{Behnke:2013lya,Gomez-Ceballos:2013zzn}.
CEPC and FCC-ee are circular colliders designed to run in some stages with center-of-mass energy from 90 GeV to 350 GeV.
ILC is a linear particle accelerator designed to run in some stages awith center-of-mass energy from 250 GeV to 1 TeV.
Due to the same types of interaction, there are similar physics processes in these colliders.
No sepcific requirements about detector geometry are applied, which minimizes the reliance of the study.
The systematic uncertainty of CEPC is supposed to be very small compared to the conserved assumption of 5\% systematic uncertainty in this paper~\cite{smiljanic:2020sys}. 
The results can be used as a guide for the similar searches on the other two colliders.

\begin{figure}[!htb]
\centering
  \subfigure [direct stau production] {\includegraphics[width=.23\textwidth]{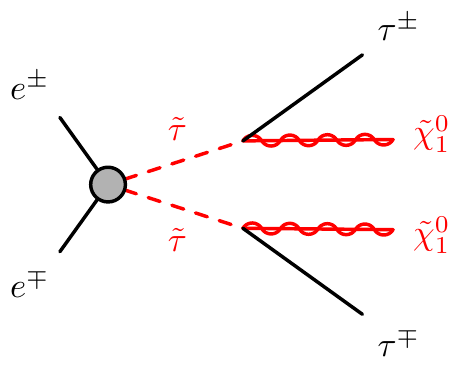}}
  \subfigure [direct smuon production] {\includegraphics[width=.23\textwidth]{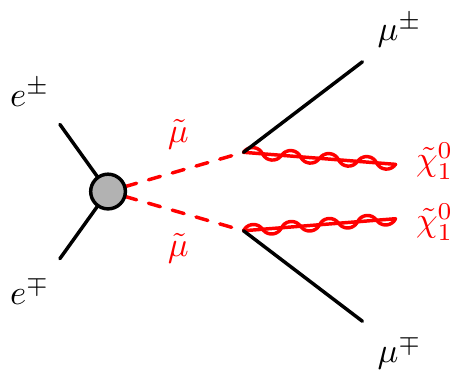}}
\caption{Representative diagram illustrating the pair production of charged staus (smuons) and subsequent decay into a two-tau (two-muon) final state}
  \label{fig:feynmanslepton}
\end{figure}

This paper presents the sensitivity studies of the direct stau / smuon production as illustrated in Figure \ref{fig:feynmanslepton}.

%% file: tex/samples_software.tex
The CEPC Conceptual Design Report (CDR) presents the comprehensive introduction of detector and software~\cite{CEPCStudyGroup:2018ghi}.
There are two types of CEPC detector suggested.
The baseline detector follows particle flow principle, and uses an ultra high granularity calorimetry system, a low material silicon tracker and a 3 Tesla magnitude filed.
The alternative detector uses 2 Tesla solenoid, a dual readout calorimeter and a drift chamber.
In this study, the baseline detector is used in the MC simulation.

The software used in the process of simulation is as follows:
The official SM background samples is generated by Whizard~\cite{Kilian:2011sepTEPJC}.
The SUSY signal samples are generated using MadGraph~\cite{Alwall:2014julJHEP} and Pythia~\cite{Sjostrand:2014zea}.
The interactions between particles and detector material are simulated by MokkaC~\cite{MorasMokka}.
The tracks reconstruction are done by Clupatra~\cite{Gaede_2014}.
The particle flow algorithm Arbor~\cite{Ruan:2018yrh} is used to reconstruct physics object.
The LICH~\cite{Yu:2017mpx} based on Multivariate Data Analysis (TMVA)~\cite{hoecker2007tmva} is used for lepton identification.

The sleptons (stau or smuons) are pair produced after electron positron collisions and each slepton decay into a lepton and a \ninoone with a 100\% branch ratio.
In the search for direct slepton pair production, the lightest neutralino is the LSP and purely Bino.
All sparticles except those mentioned here are assumed to be inaccessible at CEPC energy.
The masses of all charginos and neutralinos apart from LSP were set to 2.5 TeV to forbid all other decay channels.
The mixing matrix for the scalar taus / muons is antidiagonal such that no mixed production modes are expected.
The signal samples of direct stau (smuon) production are parametrized as function of the \stau\ (\smu) and LSP masses, where the lower \stau\ (\smu) mass is bounded by LEP limit and the LSP mass is bounded by the \stau\ (\smu) mass. 
The masses of \stau and \smu varied in the range 80 - 119 GeV.
In this work, the superpartner of the left-handed lepton and right-handed lepton are considered to be mass degenerate.
For each signal point, $10^5$ events are simulated.
Reference points with \stau ( \smu\ ) masses of 115 GeV and \ninoone mass of 20, 60, 100 ( 20, 70, 110 ) GeV are used in this paper to illustrate typical features of the SUSY models to which this analysis is sensitive. The theoretical cross sections calculated by MadGraph~\cite{Alwall:2014julJHEP} at leading order (LO) was 23.6 fb for \stau and \smu with mass of 115 GeV.

In this study, only the SM processes with two leptons (electrons, muons or taus) and large recoil mass are taken into account.
The background processes includes two fermions processes, four fermions processes and Higgs signal.
The Higgs signal is $\nu\nu H, H\to\tau\tau$.
The two fermions processes are $\mu\mu$ and $\tau\tau$ processes. 
The four fermions processes compose of ZZ,  WW, single Z, single W and Z or W mixxing processes. 
The samples are normalized to the integrated luminosity of 5.05 ab$^{-1}$.

%% file: tex/searchdl.tex
The following considered variables are effective to distinguish the signal from SM backgorunds:
\begin{itemize}
\item $|\Delta \phi(\ell^{\pm},recoil)|$, the difference of azimuth between one lepton and the recoil system.
\item $|\Delta \phi(\ell,\ell)|$, the difference of azimuth between two leptons. 
\item $\Delta R(\ell^{\pm},recoil)$, the cone size between one lepton and the recoil system.
\item $\Delta R(\ell,\ell)$, the cone size between two leptons.
\item $E_{\ell^{\pm}}$, the energy of one lepton.
\item sum$P_T$, the sum of the tranverse momentum of two leptons.
\item $M_{\ell\ell}$, the invariant mass of two leptons.
\item $M_{recoil}$, the invariant mass of the recoil system.
\end{itemize}

The signal regions are defined using the above kinematics selection criteria. Zn ~\cite{Cowan:2012ds} was used as a sensitivity reference as shown in formula (\ref{equ:Zn}). The statistical uncertainty and 5\% flat systematic uncertainty are considered in the Zn calculation.
\begin{small}
\begin{equation}
  Zn=\left[2\left((s+b)\ln\left[\frac{(s+b)(b+\sigma_b^2)}{b^2+(s+b)\sigma_b^2}\right]-\frac{b^2}{\sigma_b^2}\ln\left[1+\frac{\sigma_b^2s}{b(b+\sigma_b^2)}\right]\right)\right]^{1/2} \label{equ:Zn}
\end{equation}
\end{small}

\subsection{Search for the direct stau production}
\label{subsec:searchdt}
\input{tex/searchdt.tex}

\subsection{Search for the direct smuon production}
\label{subsec:searchdm}
\input{tex/searchdm.tex}

\subsection{Summary of slepton search}
\label{subsec:sumsl}
\begin{figure}[!htb]
\centering
\includegraphics[width=.3\textwidth]{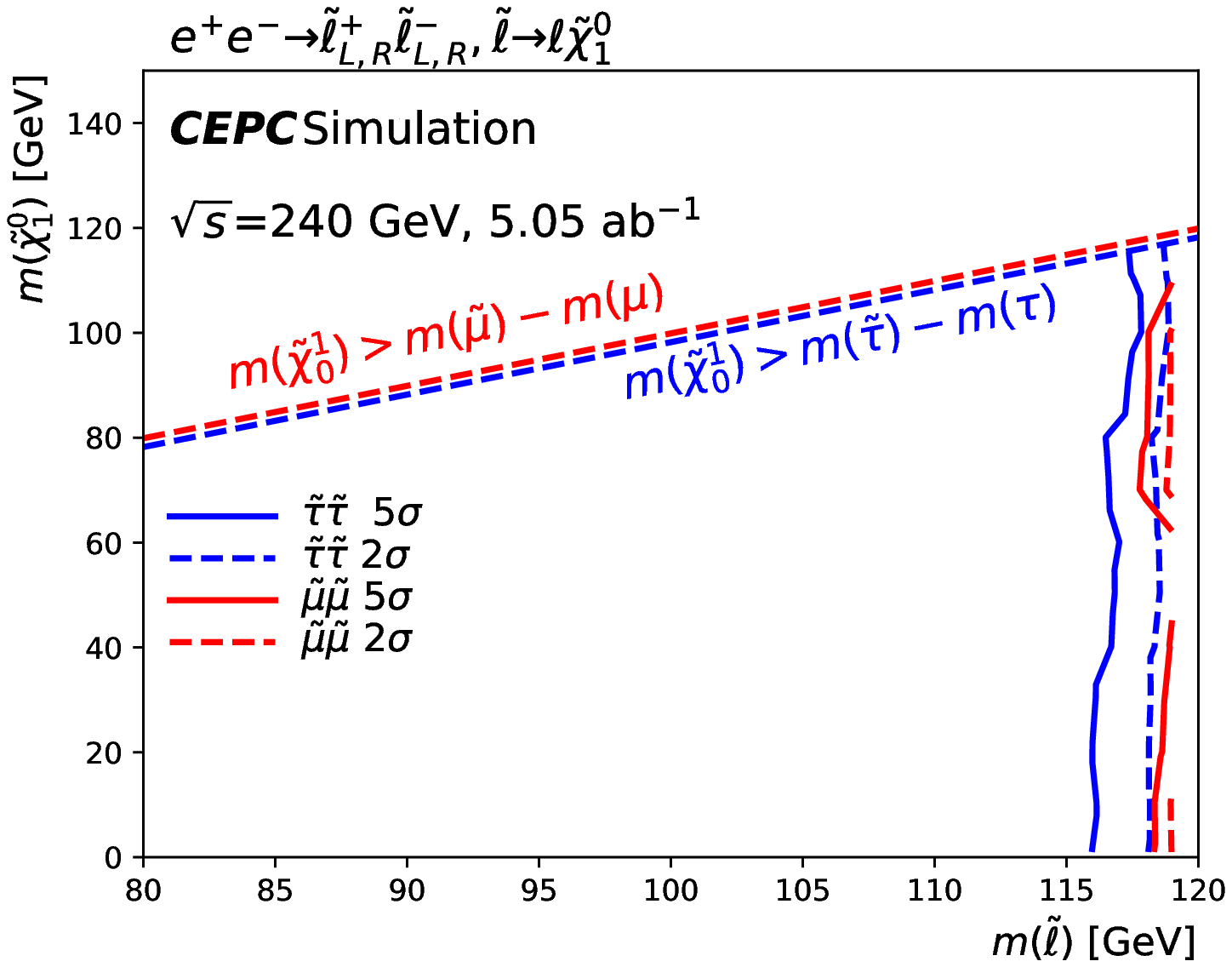}
\caption{The 5 $\sigma$ contour of direct \stau production and direct \smu production with 5\% flat systematic uncertainty.}
  \label{fig:sum}
\end{figure}

The 5 $\sigma$ contours with 5\% flat systematic uncertainty of these two scenarios are shown in Figure \ref{fig:sum}.
With the assumption of 5.05 ab$^{-1}$ and 5\% flat systematic uncertainty, the discovery potential can reach up to 116 GeV(117 GeV) for direct \stau (\smu) production.

%% file: tex/searchdt.tex
In the search for the direct stau production, the leading track with negative (positive) charge to represent the tau (anti-tau) lepton.
Events with 2 OS $\tau$ with energy \textgreater 0.5 GeV are selected.
The upper cuts on $E_{\tau^{\pm}}$ are used to suppress $\mu\mu$, Z or W mixing, WW, We$\nu$ and Ze processes. 
The sum$P_{T}$ selections are required to suppress $\tau\tau$, $\mu\mu$, Z$\nu$ processes.
The $\tau\tau$ and $\mu\mu$ processes can further be suppressed by $|\Delta \phi(\tau,\tau)|$ and $|\Delta \phi(\tau^{\pm},recoil)|$ selections.
The processes of $\tau\tau$, ZZ and single Z are rejected by $\Delta R(\tau,\tau)$ and $\Delta R(\tau^{\pm},recoil)$ selections.
The upper cuts on $M_{\tau\tau}$ are used to suppress $\tau\tau$ and We$\nu$ processes.
According to the signal topology, most of the signal events have large recoil mass. 
So, a lower cut of the invariant mass of recoil system, $M_{recoil}$, has been used to reject $\tau\tau$ and We$\nu$ processes and some other SM processes without large recoil mass.
The signal regions have been defined in Table \ref{tab:SRdt}.
Due to different behaviors of signal processes with different mass splitting between \stau and \ninoone , three signal regions are defined to cover the whole \stau-\ninoone mass parameter space well.
The SR-highDeltaM covers the region with high mass split between \stau and \ninoone, the SR-midDeltaM covers the region with medium mass split between \stau and \ninoone, and the SR-lowDeltaM covers the region with low mass split between \stau and \ninoone .
The behaviors of left-handed stau and right-handed stau are similar, so that the same signal regions are defined for both of them.

\begin{table}[hb]
  \centering
  \caption{Summary of selection requirements for the direct stau production signal region. DeltaM means difference of mass between $\stau$ and LSP}
  \label{tab:SRdt}
\begin{tabular*}{\columnwidth}{@{\extracolsep{\fill}}ccc@{}}
\hline
 SR-highDeltaM & SR-midDeltaM & SR-lowDeltaM \\
\hline
$E_{\tau^{\pm}}$ \textless 34 GeV&\multicolumn{2}{c}{$E_{\tau^{\pm}}$ \textless 15 GeV}\\
sum$P_{T}$ \textgreater 70 GeV & sum$P_{T}$ \textgreater 40 GeV & - \\
- &0.2 \textless $|\Delta \phi(\tau,\tau)|$ \textless 1.2 & $|\Delta \phi(\tau,\tau)|$ \textgreater 0.6\\
\multicolumn{2}{c}{2.4 \textless $|\Delta \phi(\tau^{\pm},recoil)|$ \textless 3} & $|\Delta \phi(\tau^{\pm},recoil)|$ \textgreater 2.3\\
0.4 \textless $\Delta R(\tau,\tau)$ \textless 1 & 0.4 \textless $\Delta R(\tau,\tau)$ \textless 1.6 & -\\
\multicolumn{2}{c}{$\Delta R(\tau^{\pm},recoil)$ \textless 3.1} & $\Delta R(\tau^{\pm},recoil)$ \textless 2.9\\
$M_{\tau\tau}$ \textless 50 GeV &$M_{\tau\tau}$ \textless 40 GeV & $M_{\tau\tau}$ \textless 18 GeV\\
$M_{recoil}$ \textgreater 90 GeV & $M_{recoil}$ \textgreater 130 GeV & $M_{recoil}$ \textgreater 210 GeV\\
\hline
\end{tabular*}
\end{table}

\begin{figure}[!htp]
  \centering
    \subfigure [SR-highDeltaM: $M_{recoil}$] {\includegraphics[width=.23\textwidth]{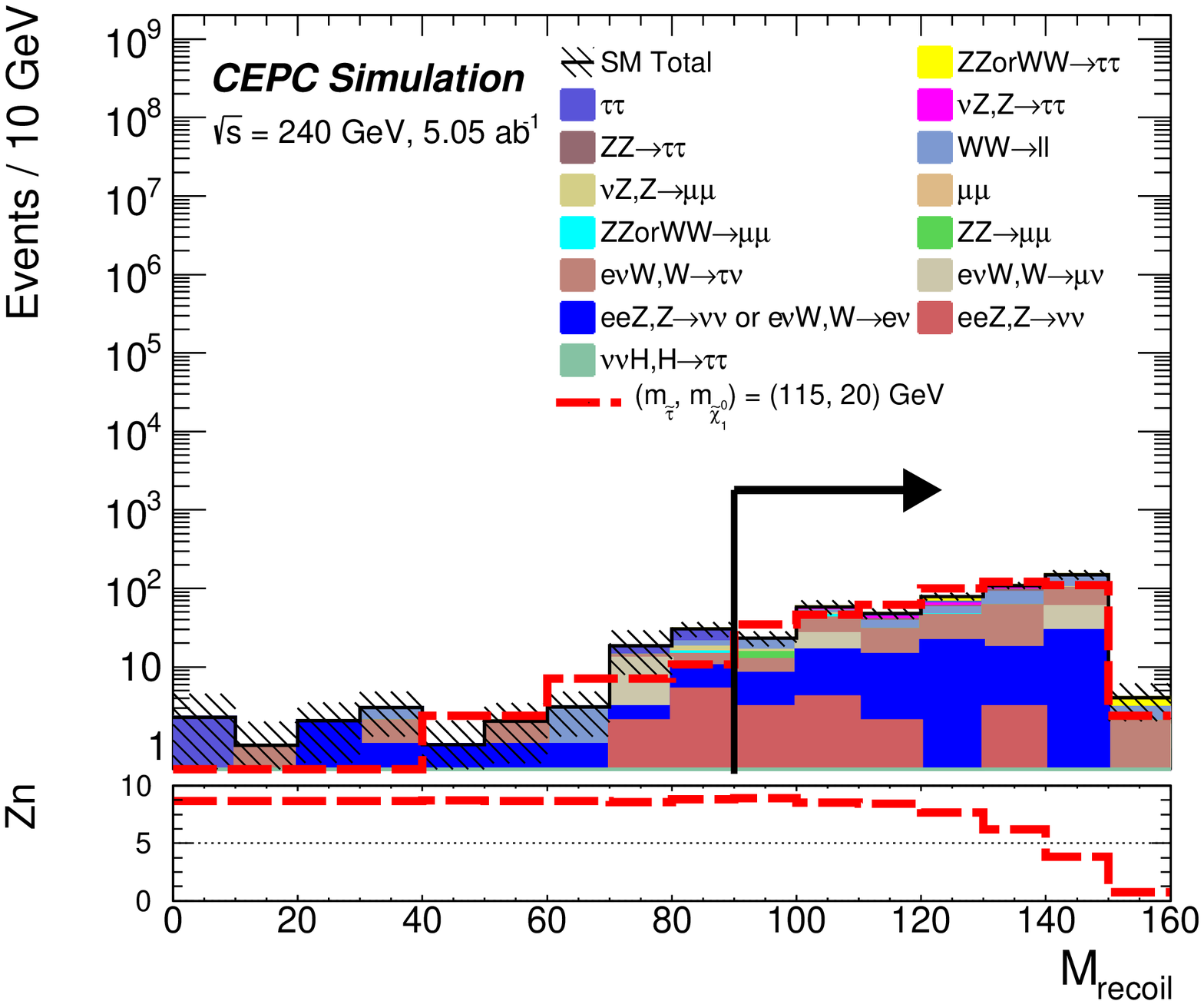}}
    \subfigure [SR-highDeltaM: $M_{\tau\tau}$]{\includegraphics[width=.23\textwidth]{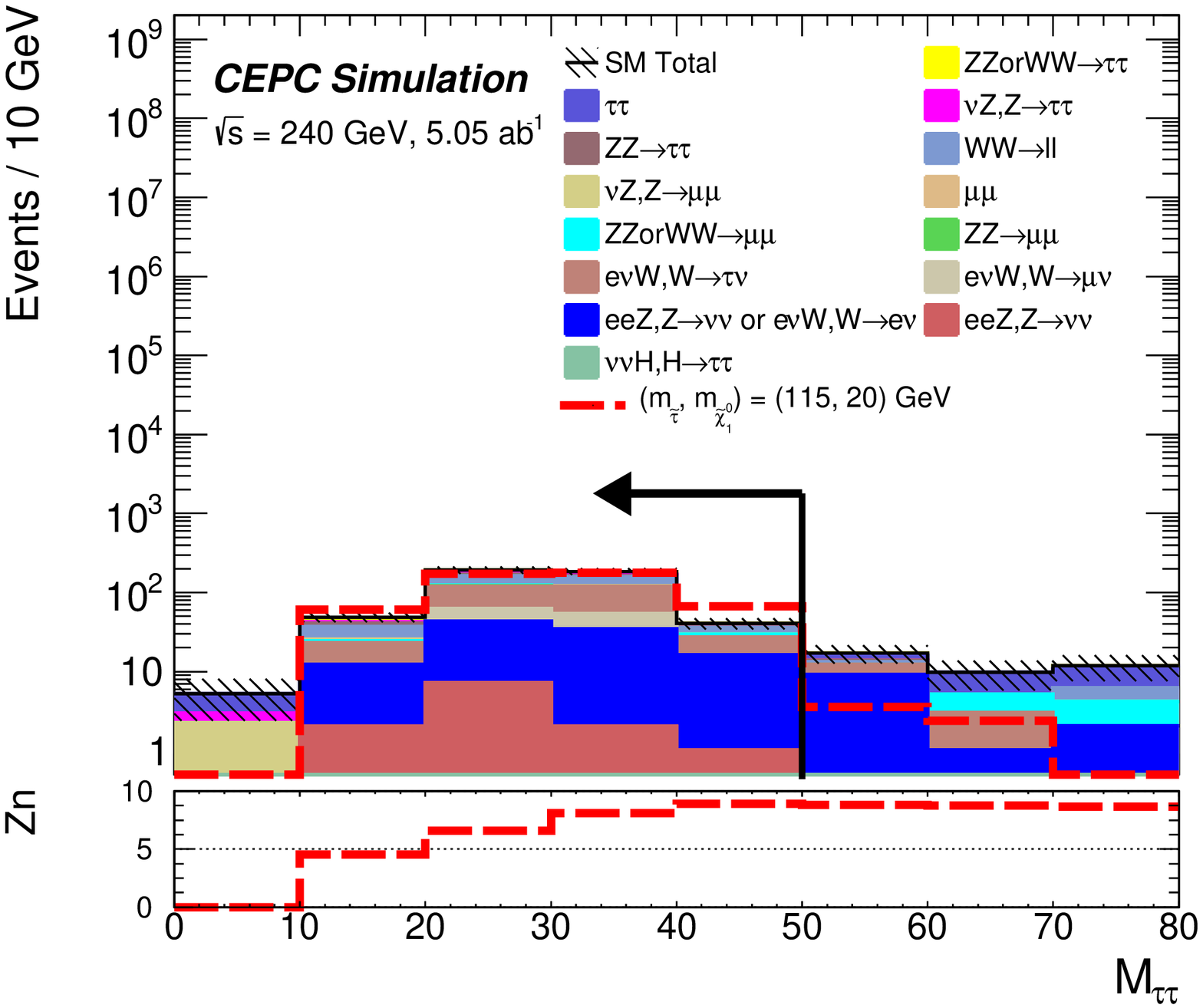}}
    \subfigure [SR-midDeltaM: $M_{recoil}$] {\includegraphics[width=.23\textwidth]{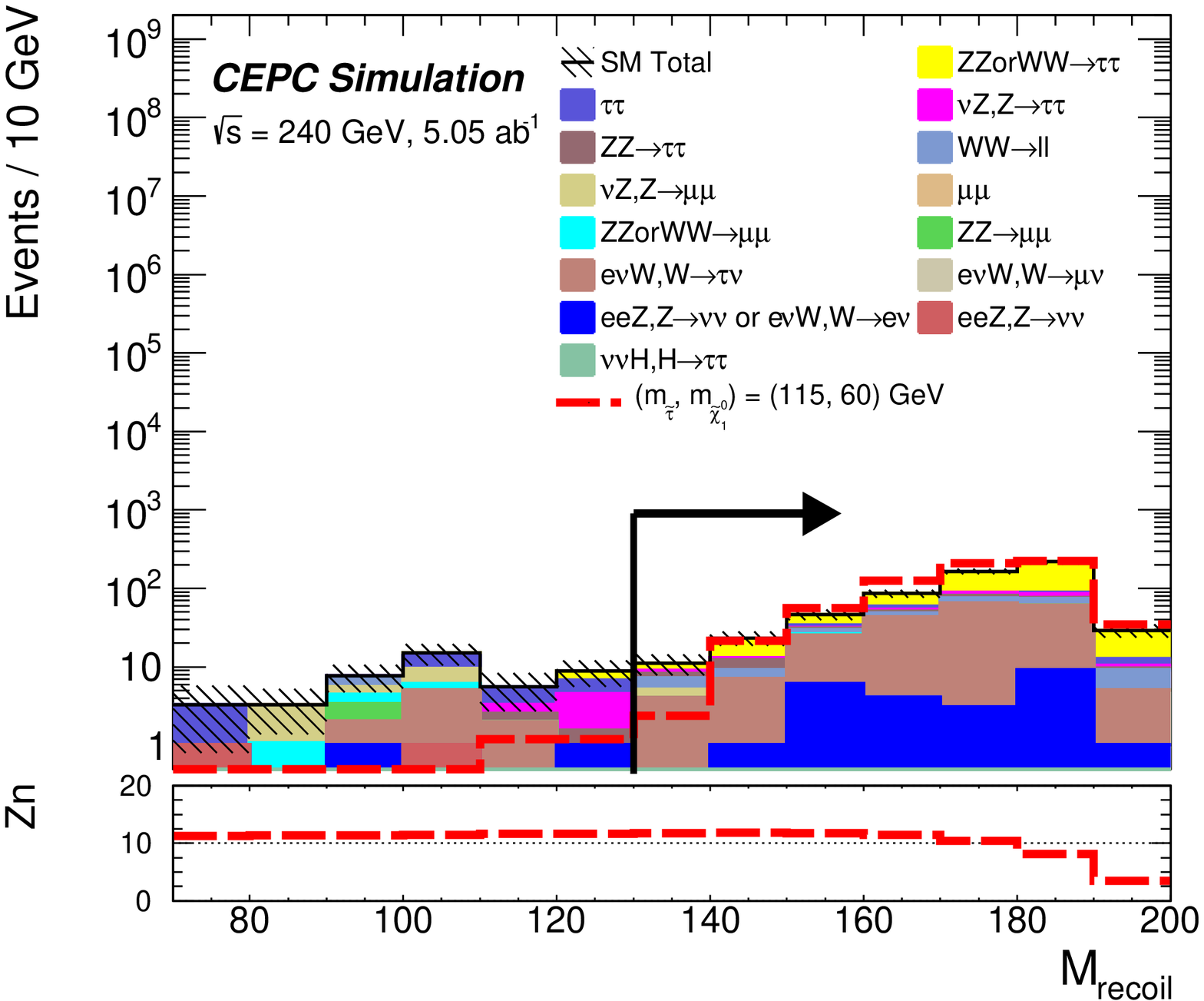}}
    \subfigure [SR-midDeltaM: $M_{\tau\tau}$] {\includegraphics[width=.23\textwidth]{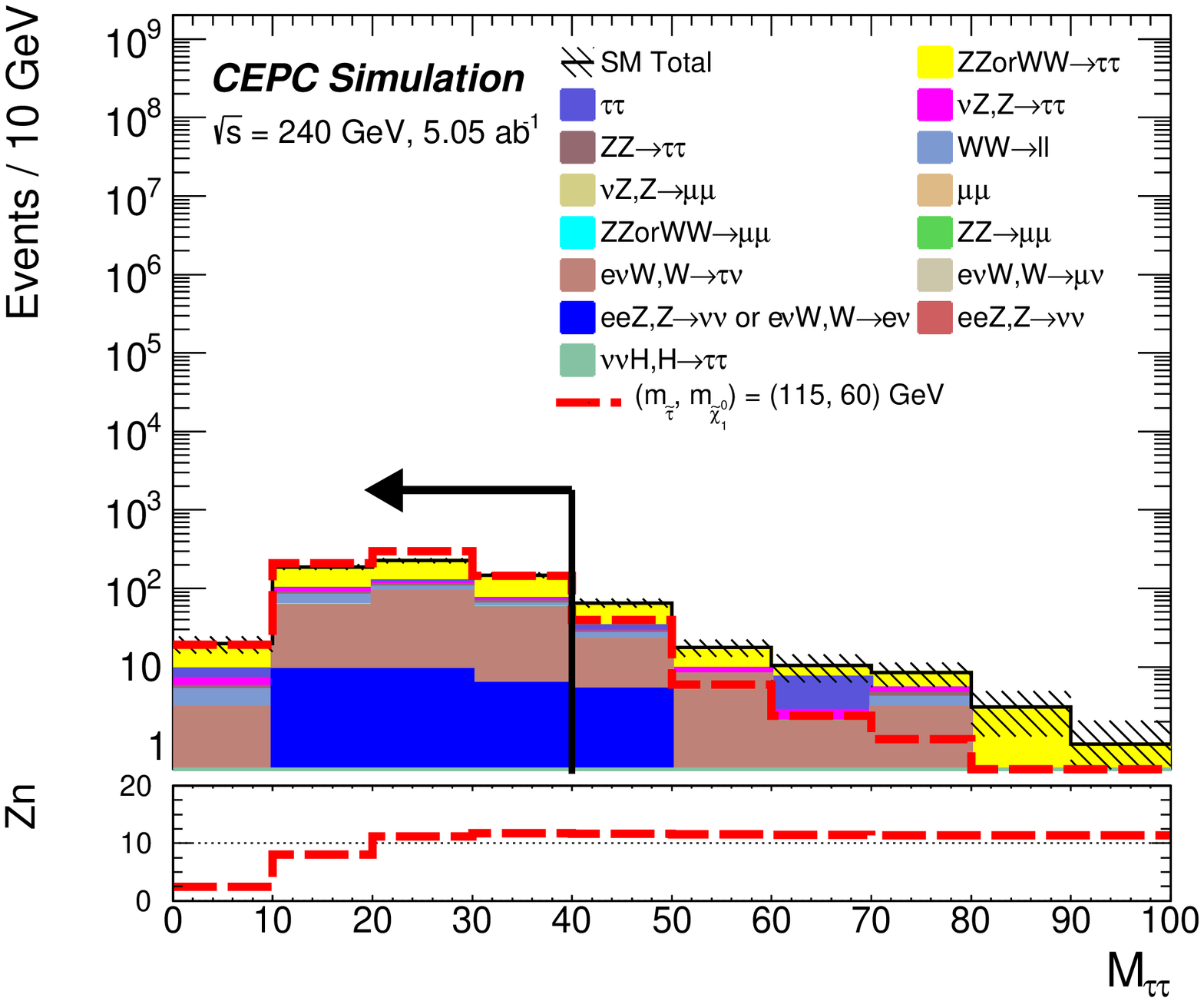}}
    \subfigure [SR-lowDeltaM: $M_{recoil}$] {\includegraphics[width=.23\textwidth]{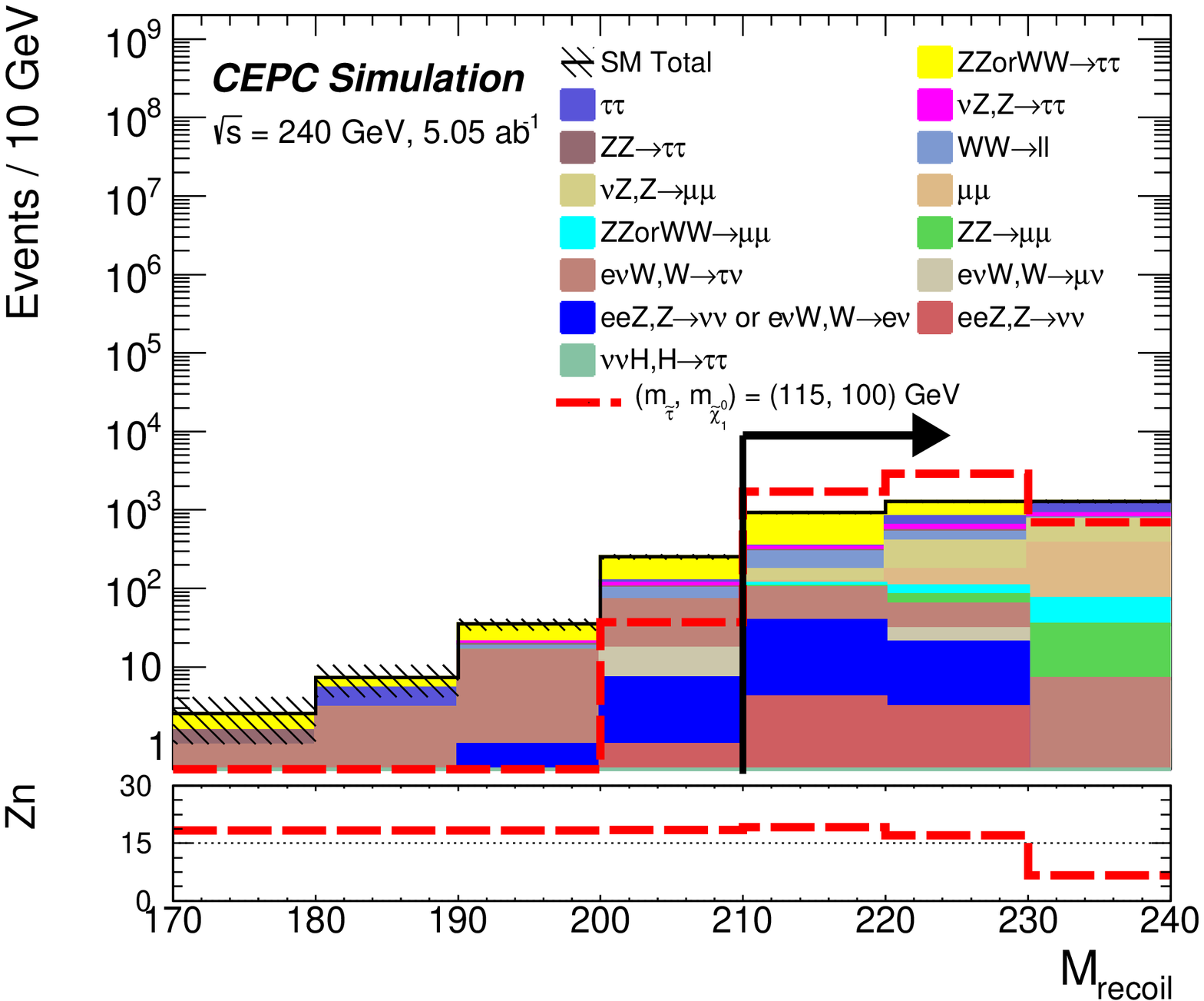}}
    \subfigure [SR-lowDeltaM: $M_{\tau\tau}$] {\includegraphics[width=.23\textwidth]{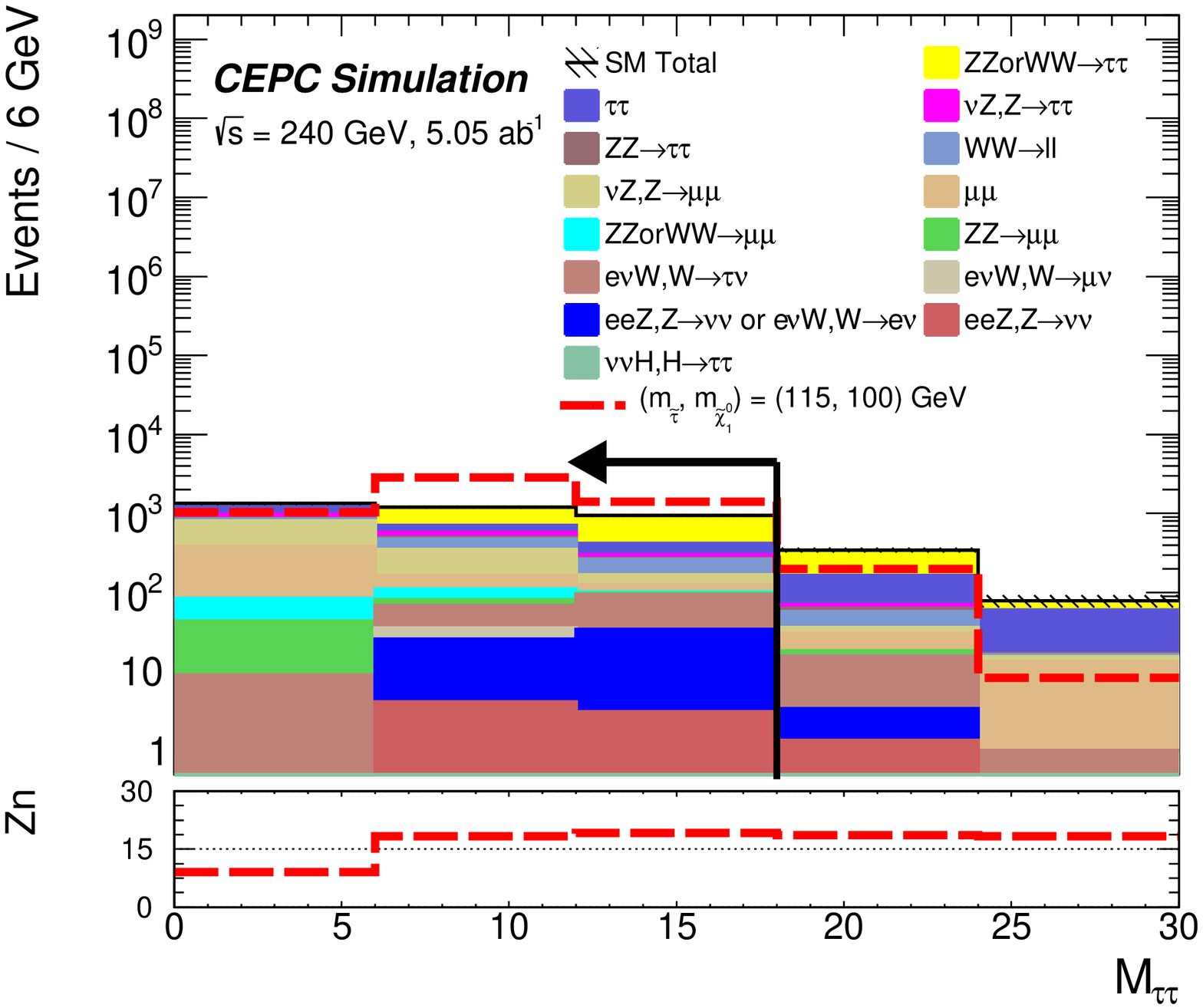}}
    \caption{"N-1" or "N" distributions of used variables after signal region requirements for direct stau production, except the variable itself, have been applied. The low pad is the Zn which calculated with statistical uncertainty and 5\% flat systematic uncertainty}
    \label{fig:nm1dt}
  \end{figure}

The kinematic distributions of $M_{recoil}$ and $M_{\tau\tau}$, after signal region requirements except the variable itself, are shown in Figure \ref{fig:nm1dt}.
The expected sensitivity Zn as function of $M_{recoil}$ and $M_{\tau\tau}$ is also shown at the lower pad of the same plot, which shows that the requirements on $M_{recoil}$ and $M_{\tau\tau}$ are efficient to distinguish between signal events and SM backgrounds events.
The event yields from the dominant background processes and the reference signal points after signal region requirements are in Table \ref{tab:numdt}, and the main background contributions are from $e\nu W, W \to \tau\nu$, $ZZ$ or $WW \to \tau\tau\nu\nu$ and $\nu Z , Z \to \mu\mu$ processes.

\begin{figure}[!htp]
  \centering
    \subfigure [systematic uncertainty = 5\%]{\includegraphics[width=.23\textwidth]{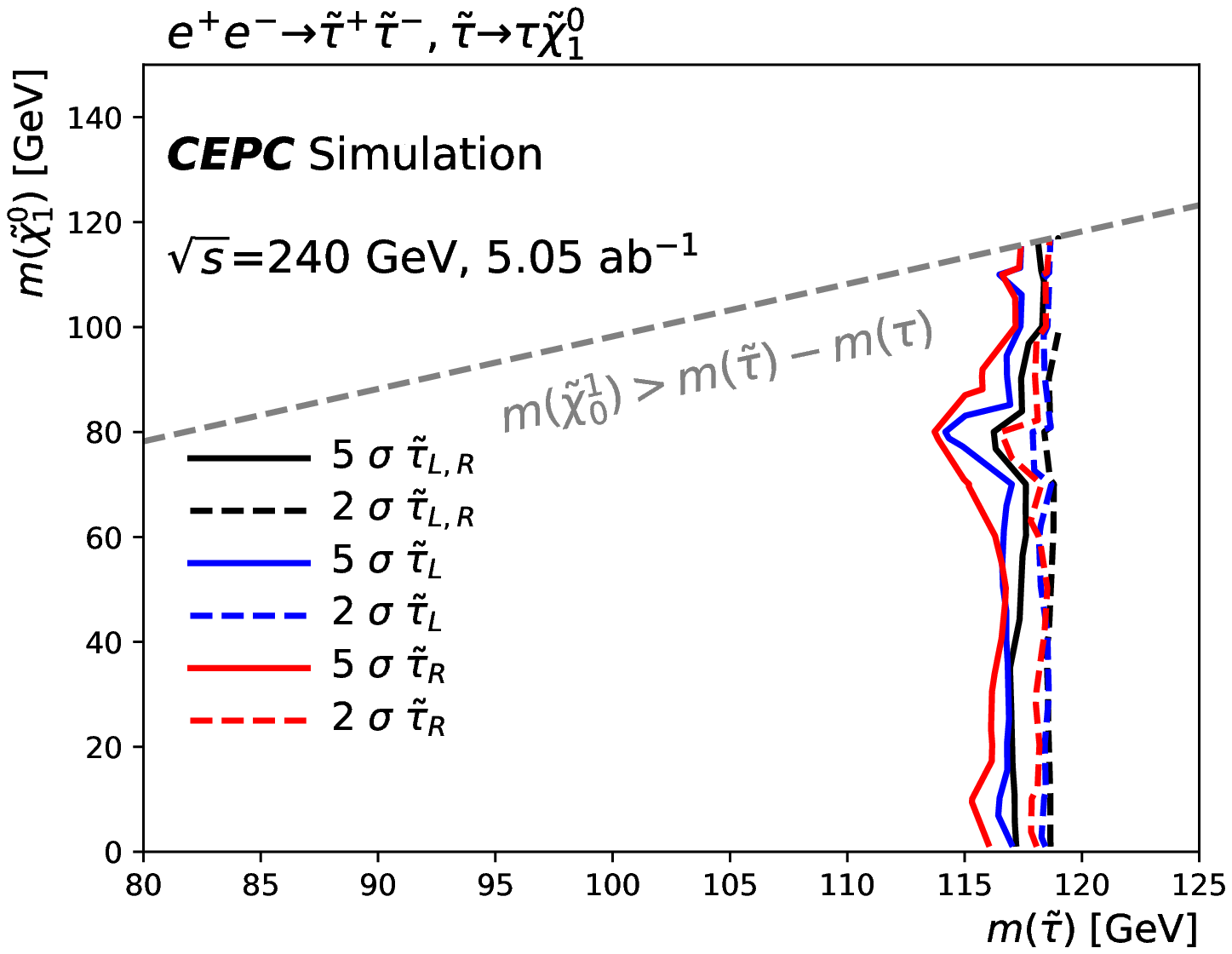}}
    \subfigure [comparison between systematic uncertainty = 0\% and 5 \%] {\includegraphics[width=.23\textwidth]{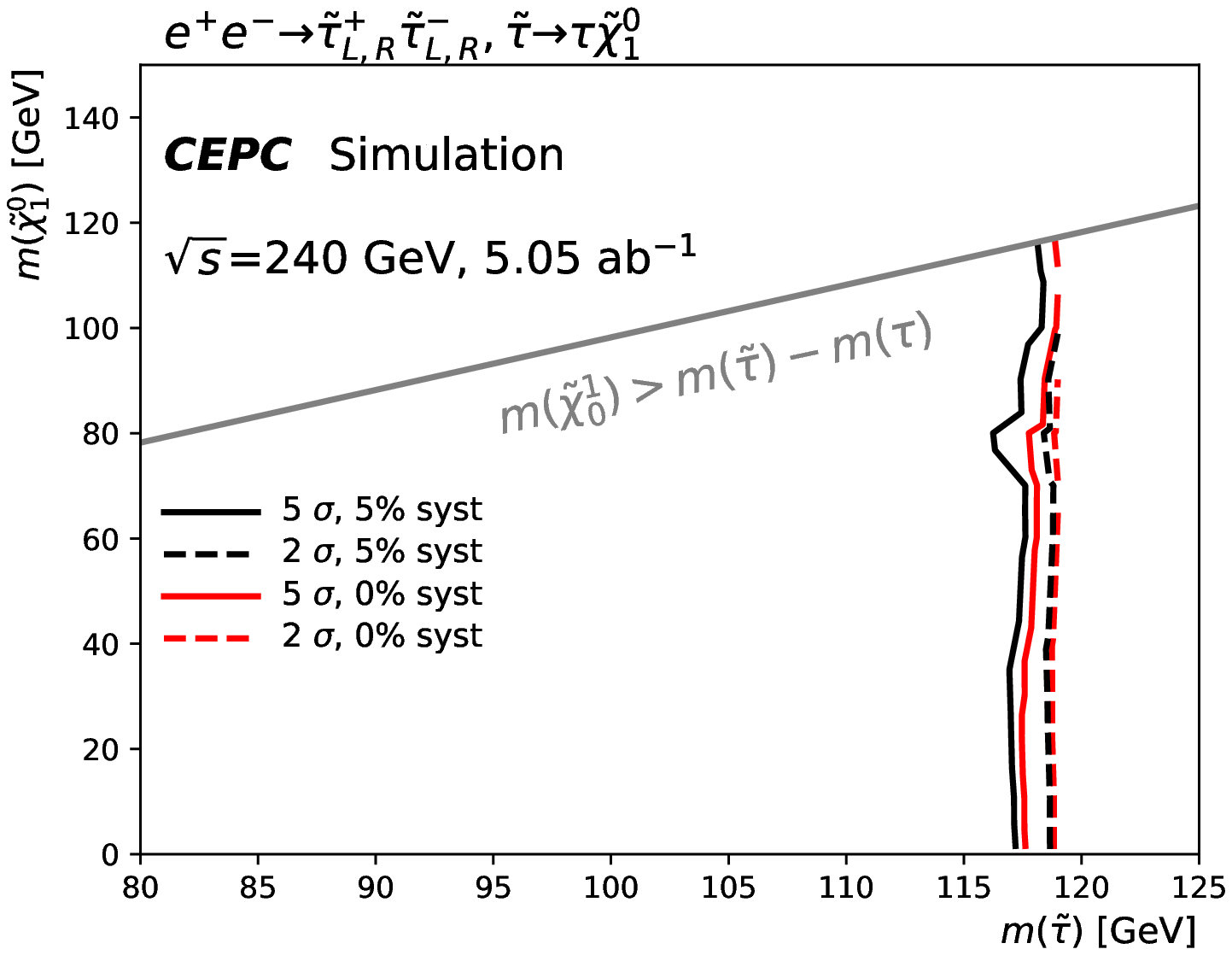}}
    \caption{The expected sensitivities as function of \stau mass and \ninoone mass for direct \stau production with systematic uncertainty of 0\% and 5\% assumption}
    \label{fig:summapdt}
\end{figure}

The expected sensitivities as function of \stau mass and \ninoone mass for the signal regions with systematic uncertainty of 0\% and 5\% for direct stau production are shown in Figure \ref{fig:summapdt}.
For each signal point, the signal region with best Zn has been chosen in sensitivity map in Figure \ref{fig:summapdt}.
With the assumption of 5\% flat systematic uncertainty, the discovery potential can reach up to 116 GeV ( 113 GeV ) with left-handed and right-handed stau (left or right-handed stau only), which is not much effected by systematic uncertainty of detectors. 

\begin{table*}
  \centering
  \caption{The number of events in the signal regions for signal and SM backgrounds with statistical uncertainty for direct stau production}
  \label{tab:numdt}
  \begin{tabular*}{\textwidth}{@{\extracolsep{\fill}}cccc@{}}
\hline
process & SR-highDeltaM & SR-midDeltaM & SR-lowDeltaM \\
\hline
$ZZ$ or $WW \to \tau\tau\nu\nu$             &  40.2$\pm$6.4  &    273$\pm$17&   1154$\pm$34\\
$\tau\tau$                                  & 6.8$\pm$3.9 & 13.6$\pm$5.6 &    493$\pm$33\\
$\nu Z , Z \to \tau\tau$                    & 15.6$\pm$3.4 &  22.2$\pm$4.1 &    224$\pm$13 \\
$ZZ \to \tau\tau\nu\nu$                     &  17.1$\pm$3.0  &  18.8$\pm$3.2&  47.2$\pm$5.0 \\
$WW \to \ell\ell$                           & 91.0$\pm$9.7 &  38.9$\pm$6.3 &    259$\pm$16\\
$\nu Z , Z \to \mu\mu$                      & 4.5$\pm$2.2 &   2.2$\pm$1.6&   698$\pm$28\\
$\mu\mu$                                    & - &-&  408$\pm$50\\
$ZZ$ or $WW \to \mu\mu\nu\nu$               & 5.3$\pm$2.4 & 1.1$\pm$1.1&   74.8$\pm$8.9\\
$ZZ \to \mu\mu\nu\nu$                       & 2.8$\pm$2.0 &-&  54.0$\pm$8.6\\
$e\nu W, W \to \tau\nu$                     & 143$\pm$12  &185$\pm$14 &   102$\pm$10\\
$e\nu W, W \to \mu\nu$                      & 39$\pm$20  &- &   9.8$\pm$9.9\\
$eeZ, Z \to \nu\nu$ or $e\nu W, W \to e\nu$ & 94.6$\pm$9.9 &  24.7$\pm$5.0& 52.4$\pm$7.3\\
$eeZ, Z \to \nu\nu$                         & 12.3$\pm$3.5 &-&  7.2$\pm$2.7\\
$\nu\nu H, H\to\tau\tau$                    & - &-&-\\
\hline
Total background                            & 473$\pm$29 & 580$\pm$24&   3582$\pm$80\\
\hline
m($\stau$,\ninoone) = (115,20) GeV          & 478$\pm$24  &  403$\pm$22&   255$\pm$17 \\
m($\stau$,\ninoone) = (115,60) GeV          &  93$\pm$11  &  671$\pm$28&    608$\pm$27 \\
m($\stau$,\ninoone) = (115,100) GeV         & -   &- &   5275$\pm$79\\
\hline      
\end{tabular*}
\end{table*}

%% file: tex/searchdm.tex
Events containing exactly two muons are selected, and the two muons have OS charge and energy larger than 0.5 GeV.
The $E_{\mu^{\pm}}$ selections are required to reject $\tau\tau$ and Z$\nu$ processes.
The cuts on $\Delta R(\mu^{\pm},recoil)$ are used to suppress $\tau\tau$, $\mu\mu$ and ZZ processes. 
The upper cuts on $M_{\mu\mu}$ are used to suppress WW and $\mu\mu$ processes.
According to the signal topology, most of the signal events have large recoil mass. 
So, a lower cut of the invariant mass of recoil system, $M_{recoil}$, has been used to reject $\mu\mu$ and Z or W mixing processes and some other SM processes without large recoil mass.
The signal regions have been defined in Table \ref{tab:SRdm}.
Due to different behaviors of signal processes with different mass split between \smu and \ninoone, three signal regions are defined to cover the whole \smu-\ninoone mass parameter space well.
The SR-highDeltaM covers the region with high mass split between \smu and \ninoone, the SR-midDeltaM covers the region with medium mass split between \smu and \ninoone, and the SR-lowDeltaM covers the region with low mass split between \smu and \ninoone .

\begin{table}
  \centering
  \caption{Summary of selection requirements for the direct smuon production signal region. DeltaM means difference of mass between $\smu$ and LSP}
  \label{tab:SRdm}
  \begin{tabular*}{\columnwidth}{@{\extracolsep{\fill}}ccc@{}}
\hline
SR-highDeltaM & SR-midDeltaM & SR-lowDeltaM\\
\hline
\multicolumn{3}{c}{== 2 muons (OS, both energy 	\textgreater 0.5 GeV) } \\
$E_{\mu}>40$ GeV& 9 GeV$<E_{\mu}<48$ GeV &-\\
$\Delta R(\mu,recoil) < $ 2.9 &\multicolumn{2}{c}{$1.5<\Delta R(\mu,recoil)<2.8$} \\
$M_{\mu\mu}<60$ GeV & $M_{\mu\mu}<80$ GeV & - \\
$M_{recoil}>40$ GeV & - & $M_{recoil}>220$ GeV\\
\hline
\end{tabular*}
\end{table}

\begin{figure}[!htp]
  \centering
    \subfigure [SR-highDeltaM:$M_{recoil}$] {\includegraphics[width=.23\textwidth]{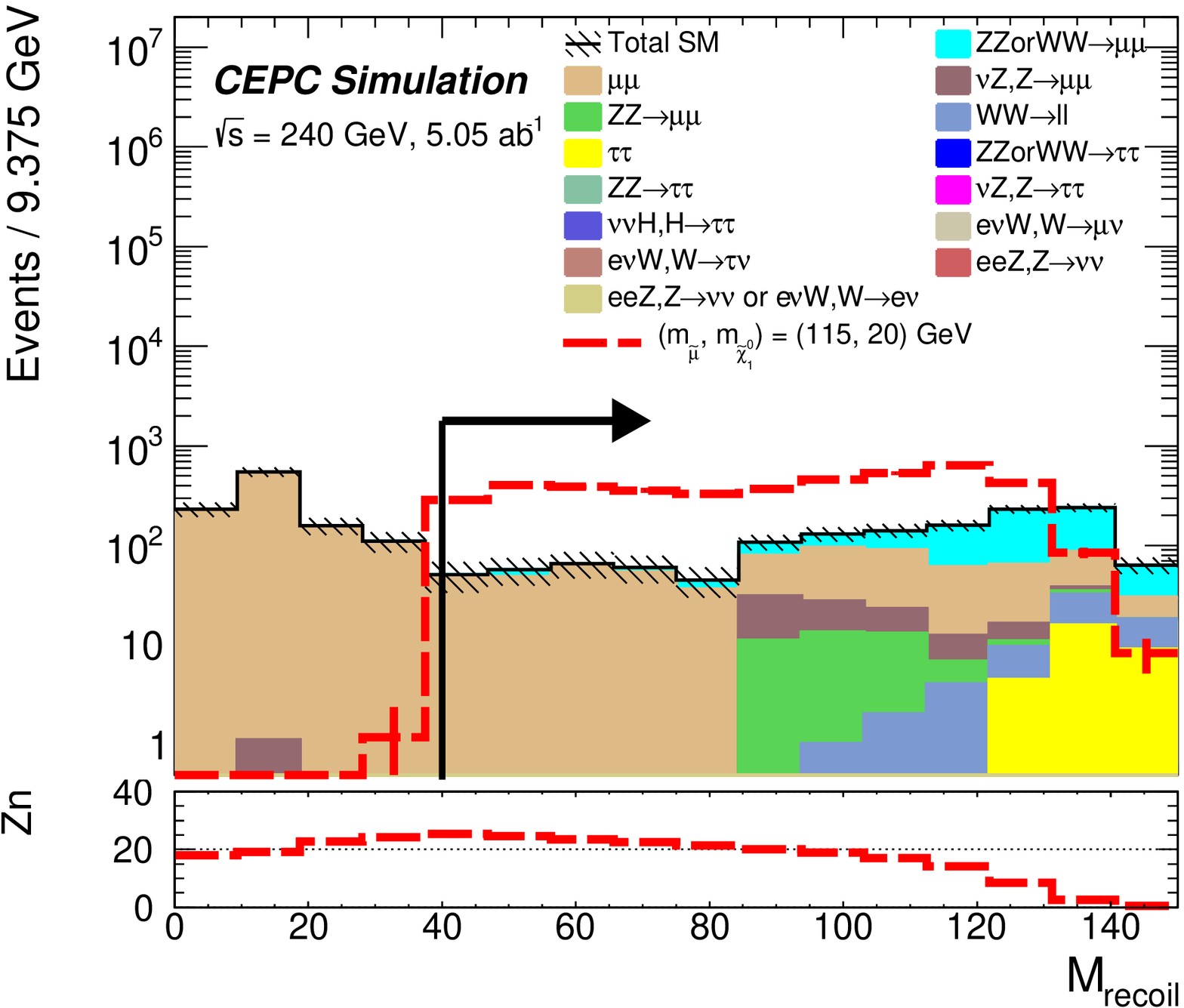}}
    \subfigure [SR-highDeltaM:$M_{\mu\mu}$] {\includegraphics[width=.23\textwidth]{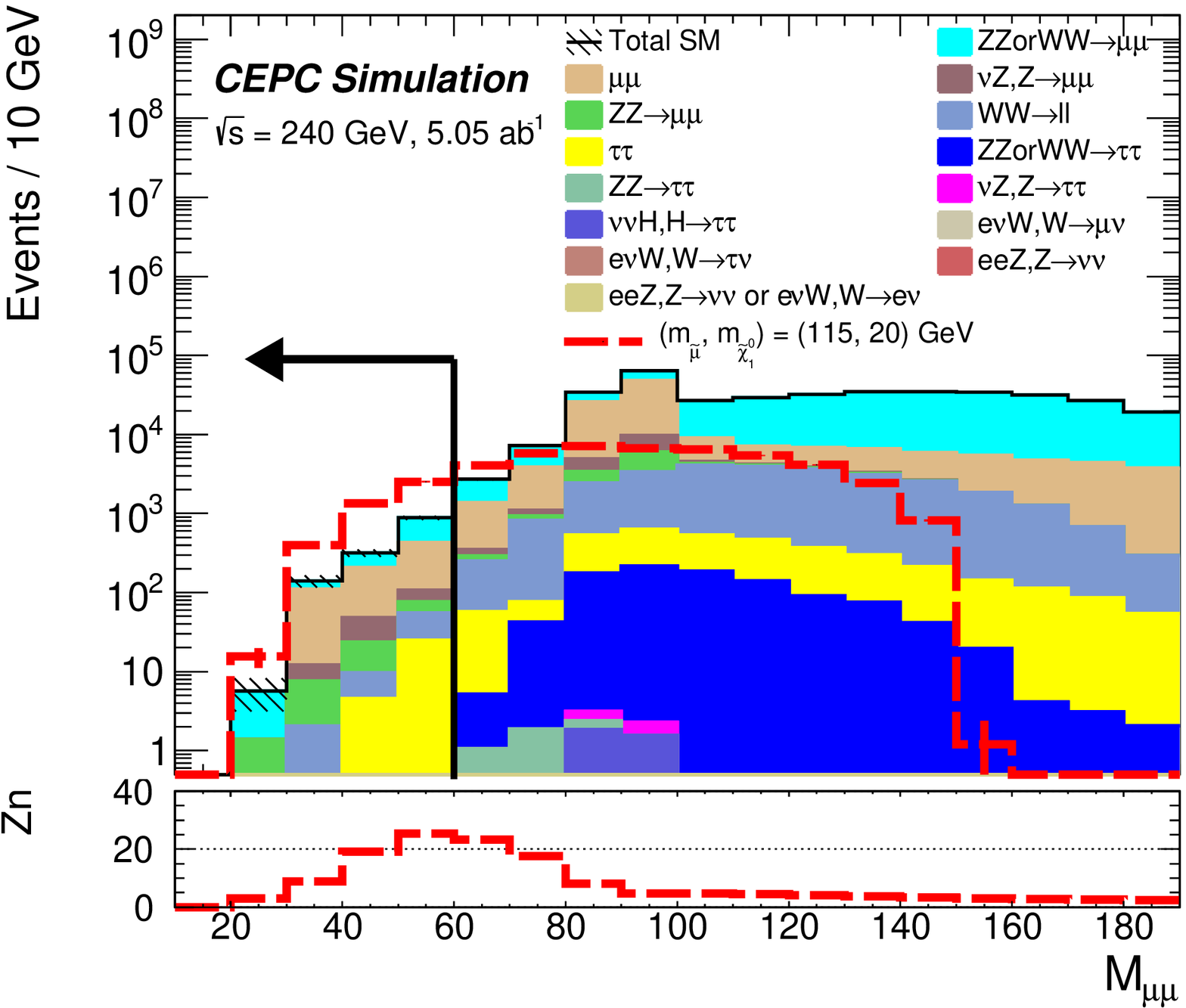}}
    \subfigure [SR-midDeltaM:$M_{recoil}$] {\includegraphics[width=.23\textwidth]{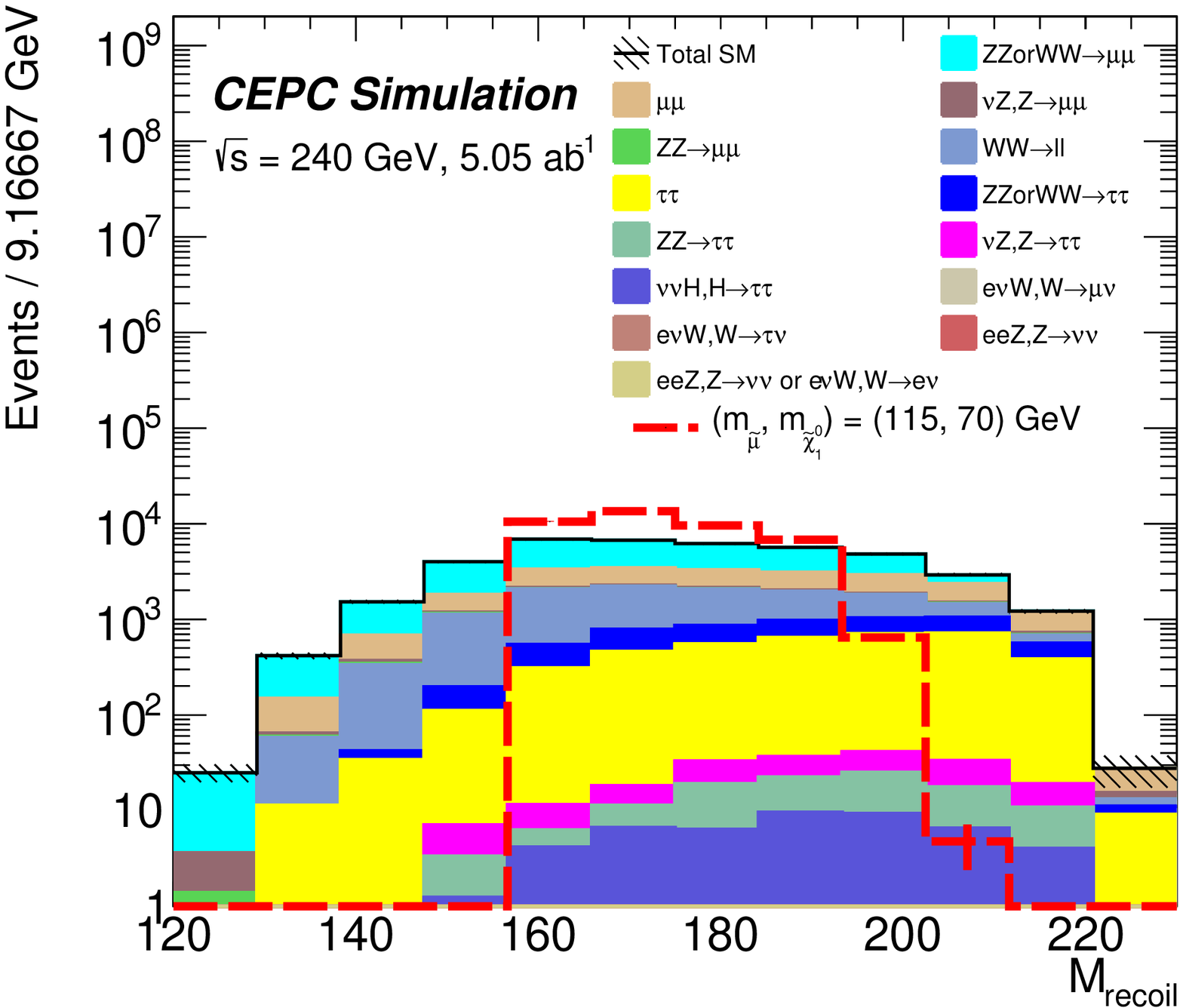}}
    \subfigure [SR-midDeltaM:$M_{\mu\mu}$] {\includegraphics[width=.23\textwidth]{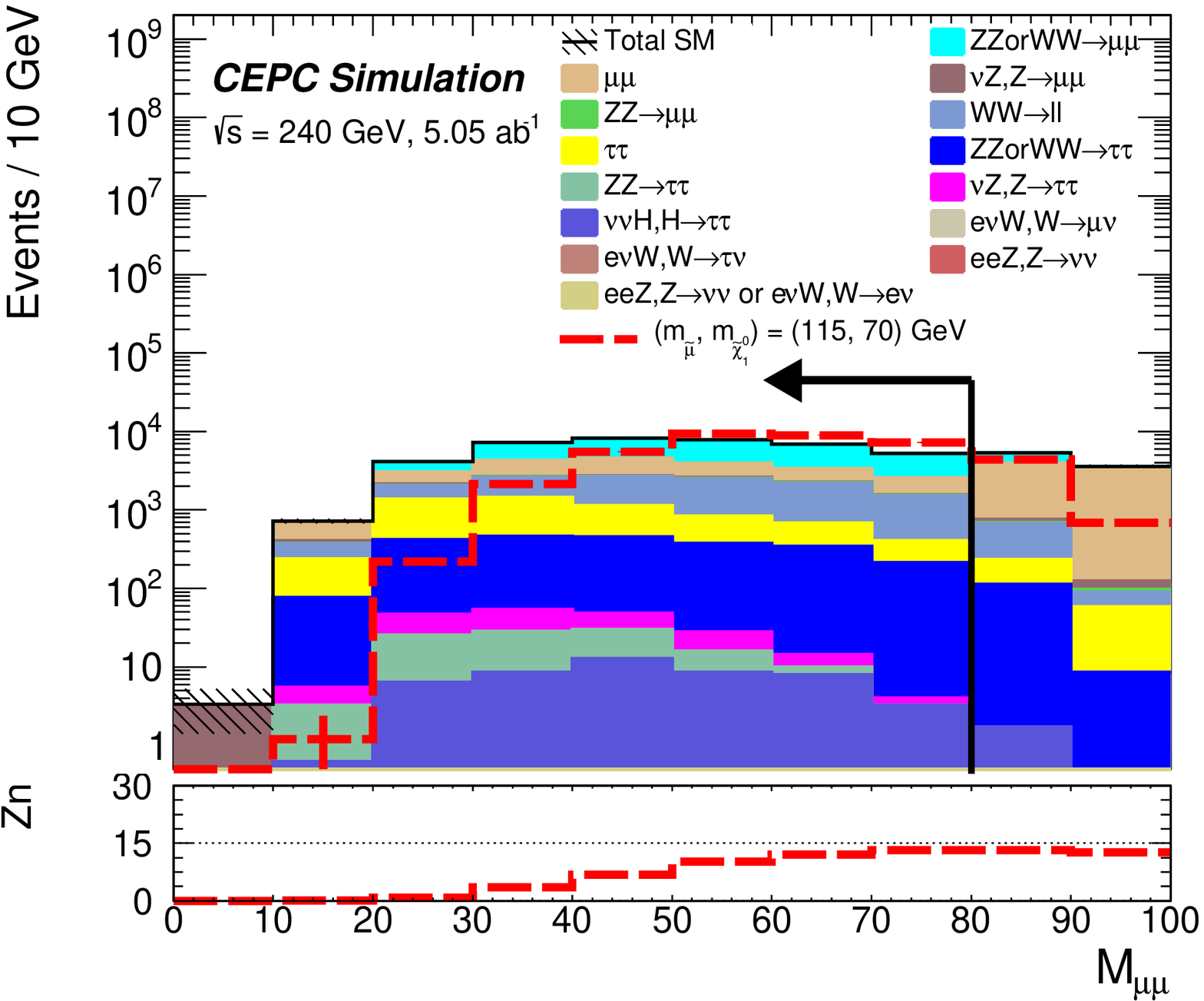}}
    \subfigure [SR-lowDeltaM:$M_{recoil}$] {\includegraphics[width=.23\textwidth]{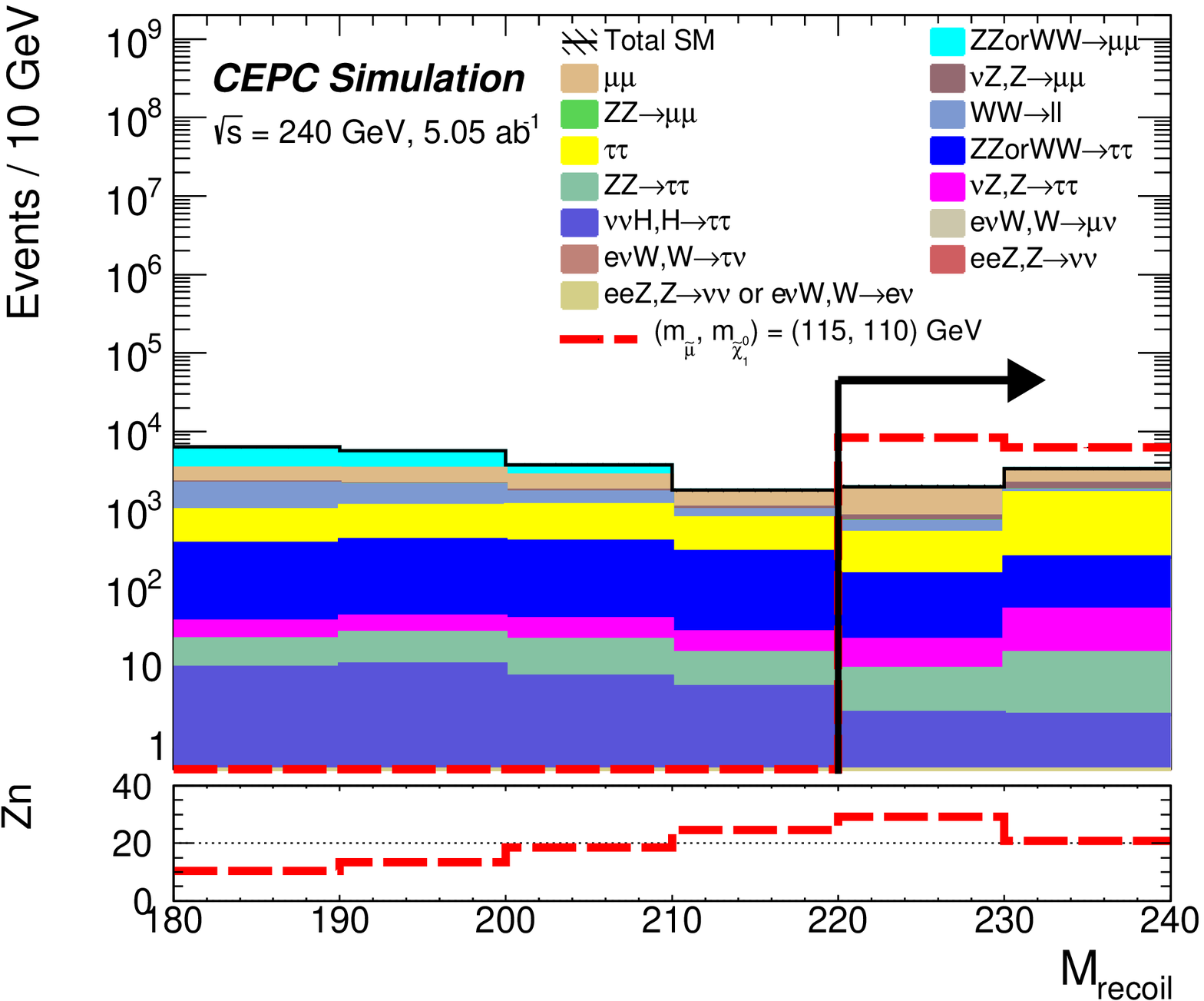}}
    \subfigure [SR-lowDeltaM:$M_{\mu\mu}$] {\includegraphics[width=.23\textwidth]{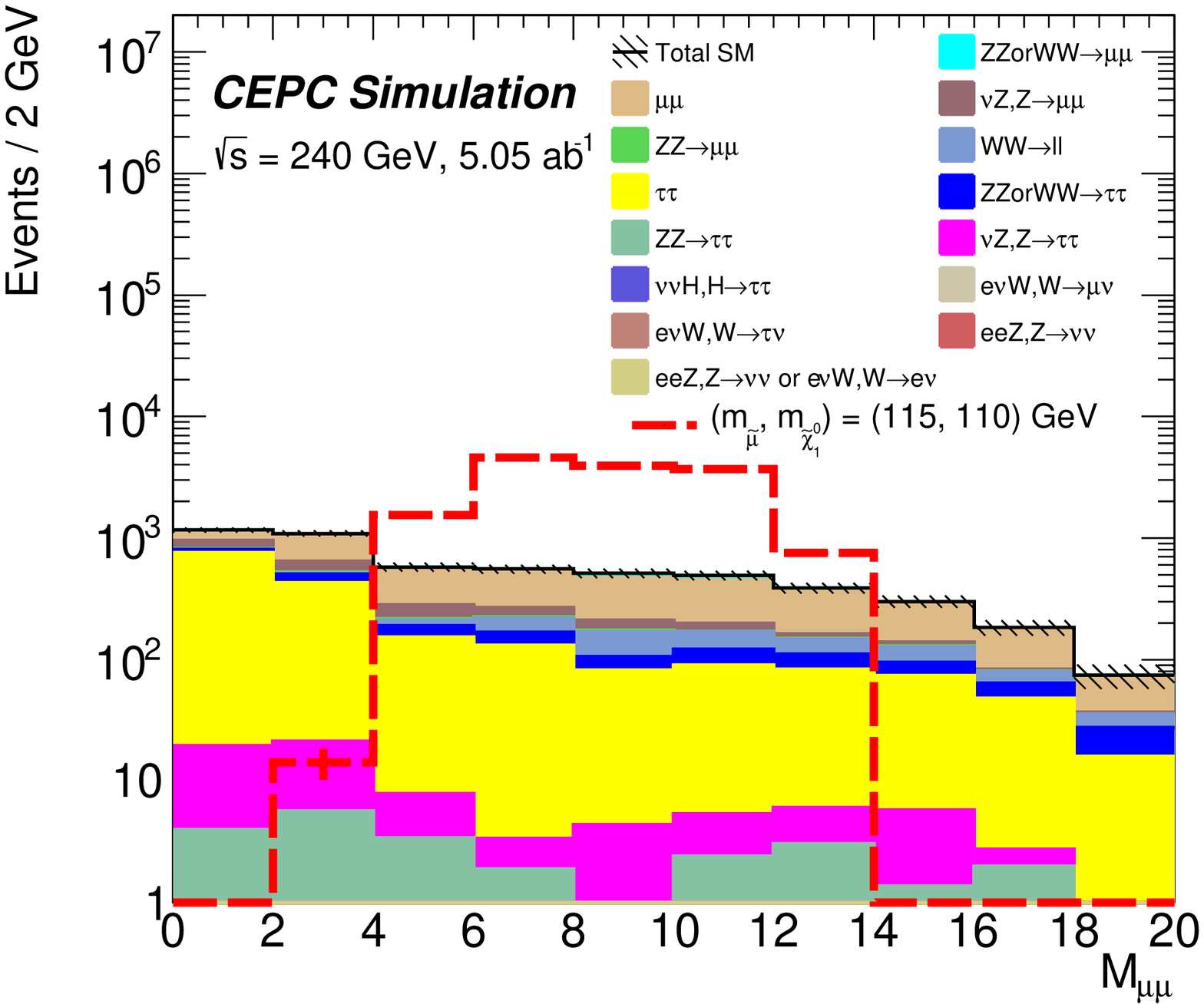}}
    \caption{"N" or "N-1" distributions of used variables after signal region requirements for direct smuon production, except the variable itself, have been applied. The low pad is the Zn which calculated with statistical uncertainty and 5\% flat systematic uncertainty}
    \label{fig:nm1dm}
  \end{figure}

\begin{figure}[!htp]
  \centering
  \subfigure [systematic uncertainty = 5\%]{\includegraphics[width=.23\textwidth]{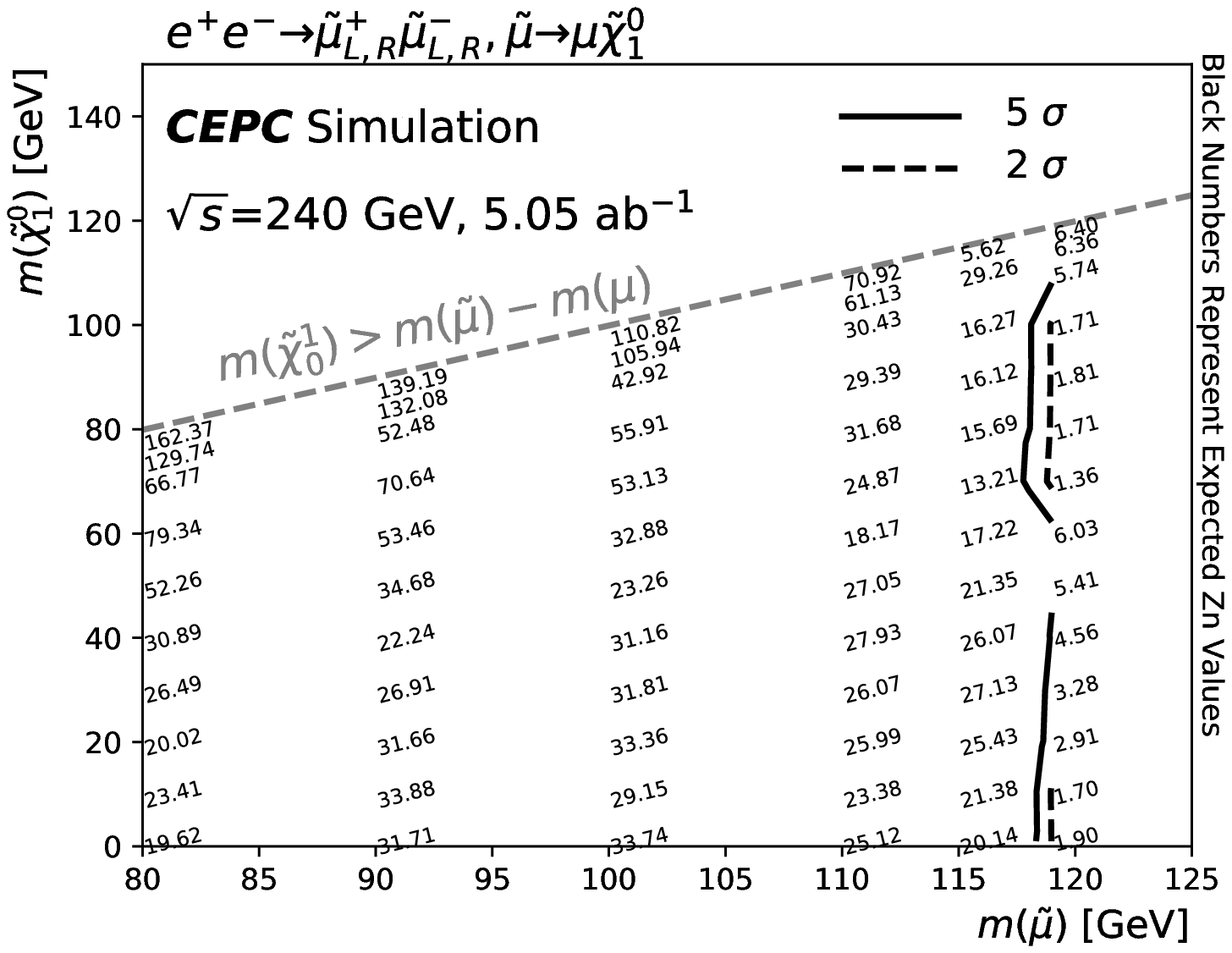}}
  \subfigure [comparison between systematic uncertainty = 0\% and 5\%] {\includegraphics[width=.23\textwidth]{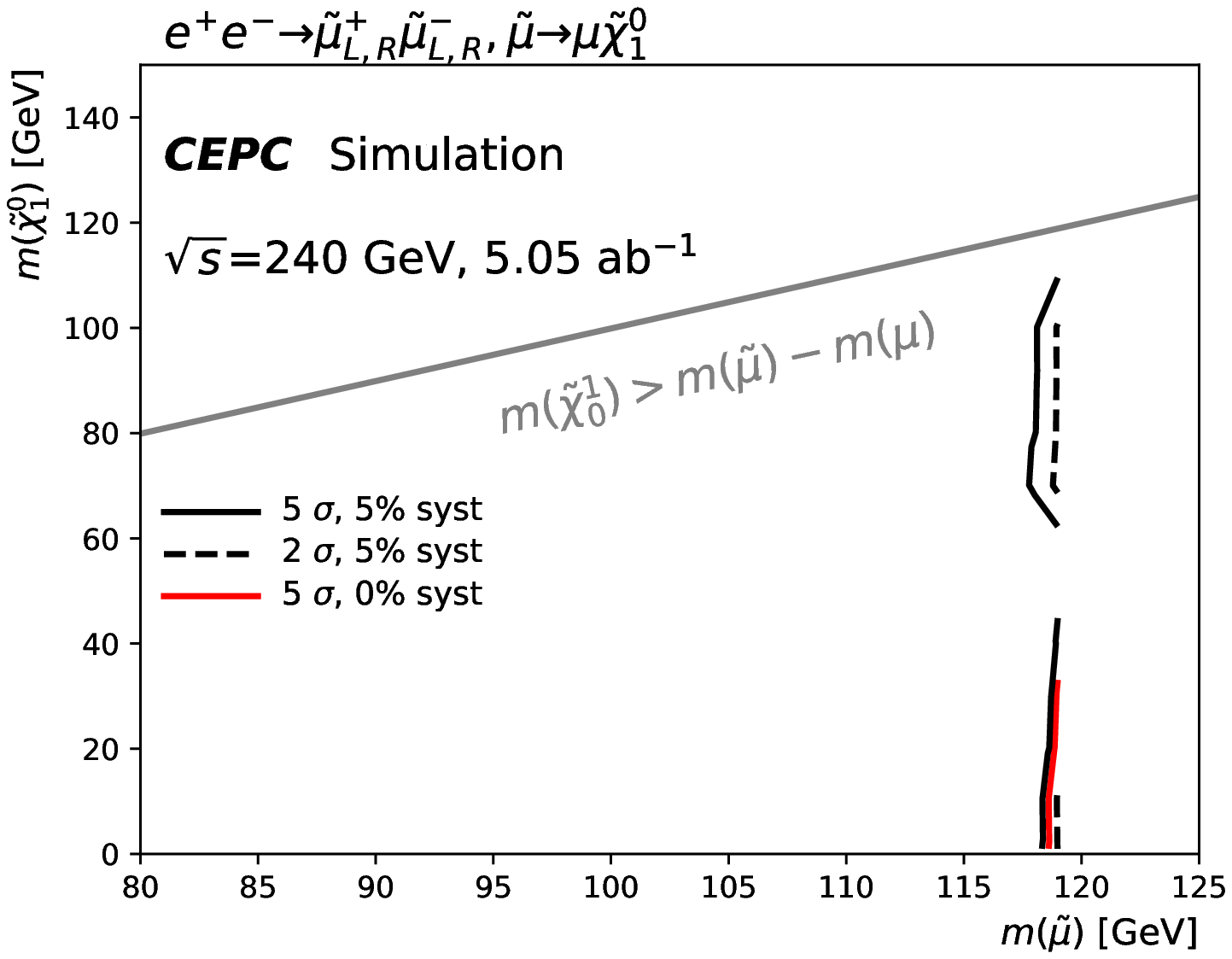}}
    \caption{The expected sensitivities as function of \smu mass and \ninoone mass for direct smuon production signal regions with systematic uncertainty of 0\% and 5\% assumption}
    \label{fig:summapdm}
\end{figure}

The kinematic distributions of $M_{recoil}$ and $M_{\mu\mu}$, after signal region requirements except the variable itself, are shown in Figure \ref{fig:nm1dm}.
The expected sensitivity Zn as function of $M_{recoil}$ and $M_{\mu\mu}$ is also shown at the lower pad of the same plot, which shows that the selections on the $M_{recoil}$ and $M_{\mu\mu}$ are efficient to distinguish between SUSY signal events and SM background processes.
The event yields from the dominant background processes and the reference signal points after signal region reqiurements are in Table \ref{tab:numdm}, and the main background contributions are from $ZZ$ or $WW \to \mu\mu\nu\nu$, $\mu\mu$ and $\tau\tau$ processes.
The expected sensitivities as function of \smu mass and \ninoone mass for the signal regions with systematic uncertainty of 0\% and 5\% for direct smuon production are shown in Figure \ref{fig:summapdm}.
For each signal point, the signal region with best Zn has been chosen in sensitivity map in Figure \ref{fig:summapdm}.
With the assumption of 5\% flat systematic uncertainty, the discovery sensitivity can reach up to 117 GeV in smuon mass, which is not too much effected by systematic uncertainty of detectors.

  \begin{table*}
    \centering
    \caption{The number of events in the signal regions for signal and SM backgrounds with statistical uncertainty for direct smuon production}
    \label{tab:numdm}
    \begin{tabular*}{\textwidth}{@{\extracolsep{\fill}}cccc@{}}
\hline
process & SR-highDeltaM & SR-midDeltaM & SR-lowDeltaM \\
\hline
$ZZorWW\to\mu\mu\nu\nu$              &     597$\pm$25 &  18020$\pm$140 &     168$\pm$13 \\
$\mu\mu$                             &    578$\pm$59 &   8000$\pm$220 &   2190$\pm$120 \\
$\nu Z,Z\to\mu\mu$                   &  59.0$\pm$8.1 &     423$\pm$22 &     467$\pm$23 \\
$ZZ\to\mu\mu\nu\nu$                  &   41.5$\pm$7.6 &     161$\pm$15 &   52.6$\pm$8.5 \\
$WW\to\ell\ell$                      &  37.9$\pm$6.2 &    7671$\pm$89 &     282$\pm$17 \\
$\tau\tau$                           &  29.5$\pm$8.2 &    3748$\pm$92 &    1782$\pm$64 \\
$ZZorWW\to\tau\tau\nu\nu$            &              - &    2128$\pm$47 &     325$\pm$18 \\
$ZZ\to\tau\tau\nu\nu$                &              - &   69.1$\pm$6.1 &   19.8$\pm$3.3 \\
$\nu Z,Z\to\tau\tau$                 &              - &   83.7$\pm$7.9 &   51.9$\pm$6.2 \\
$\nu\nu H,H\to\tau\tau$              &              - &   47.9$\pm$2.7 &  5.11$\pm$0.89 \\
$e\nu W,W\to\mu\nu$                  &               - &              - &              - \\
$e\nu W,W\to\tau\nu$                 &               - &              - &              - \\
$eeZ,Z\to\nu\nu$                     &               - &              - &              - \\
$eeZ,Z\to\nu\nu$or$e\nu W,W\to e\nu$ &               - &              - &              - \\
\hline
Total background                     &    1343$\pm$66 &  40350$\pm$300 &   5340$\pm$140 \\
\hline
m(\smu,\ninoone) = (115,20) GeV      &4288$\pm$72 &    1638$\pm$44 &              - \\
m(\smu,\ninoone) = (115,70) GeV      &             - &  41140$\pm$220 &              - \\
m(\smu,\ninoone) = (115,110) GeV     &             - &              - &  14540$\pm$130 \\
\hline 
\end{tabular*}
\end{table*}

%% file: tex/conclusion.tex
Searches for direct slepton pair production are performed at CEPC using MC simulated samples.
For direct stau production with left-handed and right-handed (left/right-handed stau only) stau, assuming flat 5\% systematic uncertainty, the discovery sensitivity can reach up to 116 GeV (113 GeV) in stau mass.
For direct smuon production, assuming flat 5\% systematic uncertainty, the discovery sensitivity can reach up to 117 GeV in smuon mass.
The \stau ( \smu\ ) mass limit extends about 30 (22) GeV beyond previous limits by LEP~\cite{LEPslepton,Heister:2001nk,Heister:2003zk,Abdallah:2003xe,Achard:2003ge,Abbiendi:2003ji} in high \stau (\smu\ ) mass region, and can cover the compressed region with small mass difference between \stau ( \smu\ ) and LSP, which is hard for ATLAS and CMS to reach~\cite{SUSY-2018-16,SUSY-2018-32,CMS-SUS-17-009,SUSY-2018-04,CMS-SUS-18-006}.
The result of this research at CEPC also can be applied to other electron-positron experiments with close center-of-mass energy, such as ILC~\cite{Behnke:2013lya} and FCC-ee~\cite{Gomez-Ceballos:2013zzn}. 

%% file: tex/acknowledgments.tex
The authors are grateful to Cheng-dong Fu, Gang Li, and Xiang-hu Zhao for providing the simulation tools.
This study was supported by the National Key Programme (Grant NO.: 2018YFA0404000).